# Empirical Thermophotovoltaic Performance Predictions and Limits

Titilope M. Dada,[1] Calvin M. Mestelle[1], Daniel J. Friedman,[3] Myles A. Steiner,[3] and Eric J. Tervo[1,2] *

[1]Department of Electrical and Computer Engineering, University of Wisconsin-Madison, Madison, WI, 53706, USA

[2]Department of Mechanical Engineering, University of Wisconsin-Madison, Madison, WI, 53706, USA

[3]National Renewable Energy Laboratory, Golden, CO, 80401, USA

*tervo@wisc.edu

## ABSTRACT

Significant progress has been made in the field of thermophotovoltaics, with efficiency recently rising to over 40% due to improvements in cell design and material quality, higher emitter temperatures, and better spectral management. However, inconsistencies in trends for efficiency with semiconductor bandgap energy across various temperatures pose challenges in predicting optimal bandgaps or expected performance for different applications. To address these issues, here we present realistic performance predictions for various types of single-junction cells over a broad range of emitter temperatures using an empirical model based on past cell measurements. Our model is validated using data from different authors with various bandgaps and emitter temperatures, and an excellent agreement is seen between the model and the experimental data. Using our model, we show that in addition to spectral losses, it is important to consider practical electrical losses associated with series resistance and cell quality to avoid overestimation of system efficiency. We also show the effect of modifying various system parameters such as bandgap, above and below-bandgap reflectance, saturation current, and series resistance on the efficiency and power density of thermophotovoltaics at different temperatures. Finally, we predict the bandgap energies for best performance over a range of emitter temperatures for different cell material qualities.

## INTRODUCTION

Thermophotovoltaics (TPVs) convert infrared photons from a hot emitter to electricity using photovoltaic cells[1–3]. As solid-state devices, they provide a number of benefits including scalability from milliwatt- to megawatt-scale systems, low maintenance and quiet operation due to the absence of moving parts, and versatility in applications such as solar-thermal energy conversion[4–7], waste heat recovery in industrial processes[8], thermal energy storage systems[9,10], and portable power generation[11], amongst others. Fig. 1a shows a schematic diagram of a TPV system. The system comprises an emitter thermally radiating towards a TPV cell connected to an external load. The performance of this system is determined by parameters such as the view factor from the cell to the emitter (how much of the cell's 'field of view' is occupied by the emitter; henceforth we will refer to this simply as the view factor), series resistance $R_{series}$ (inherent electrical resistance within the components and connections of the TPV cell), the spectral distribution of radiation energy $E$ relative to the bandgap energy $E_g$ of the TPV cell, the reflectance of the TPV cell both above and below the bandgap energy (related to the emittance $\varepsilon_{abv}$ and $\varepsilon_{bel}$), the internal quantum efficiency $IQE$ at each wavelength (a measure of the charge carrier collection efficiency), and the fraction of the minority carrier recombination which is radiative (which is related to the saturation current $J_0$). These parameters in turn lead to the electrical current-voltage characteristics under illumination, illustrated in Fig. 1b, and associated performance metrics including open-circuit voltage, short-circuit current, and fill factor.

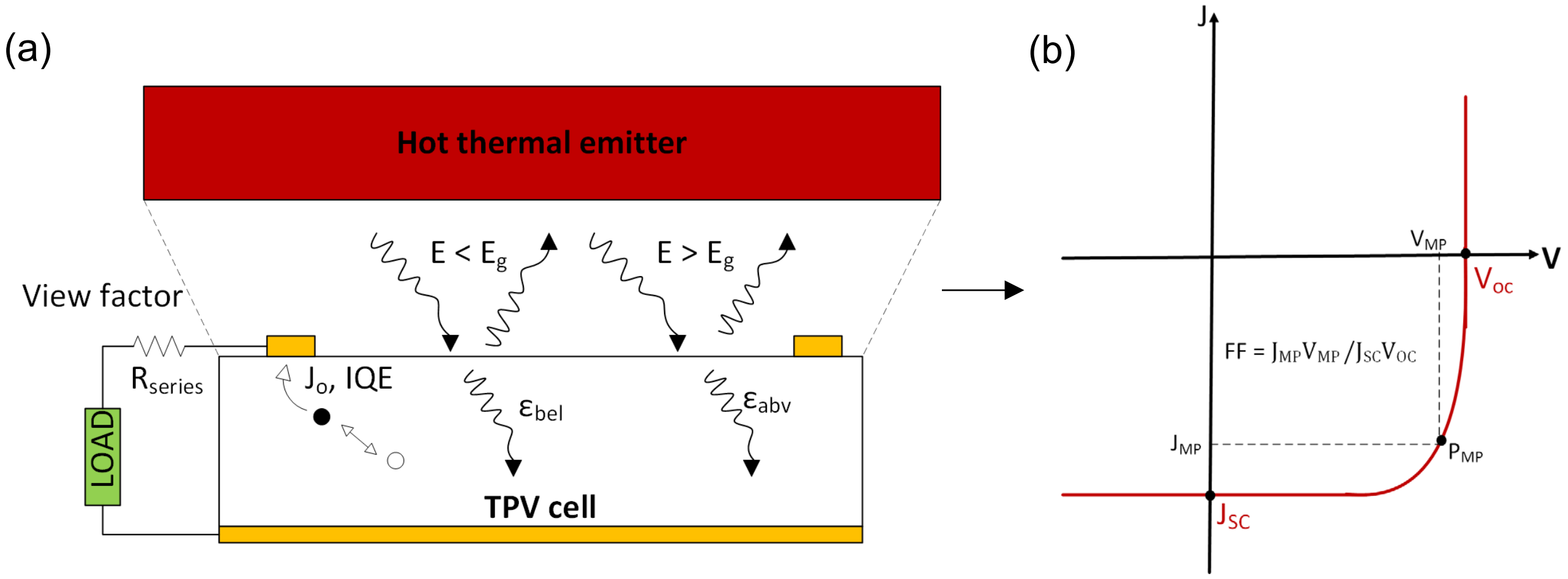

**Figure 1.** (a) Schematic of a TPV system including several parameters that affect device performance. (b) Current-voltage characteristics of a TPV cell under illumination including some important performance metrics.

Over the years, there has been amazing progress in the efficiency of TPV cells[3,12,13], defined more completely in the next section. The historical best measured efficiencies across cell materials are shown in Fig. 2. These improvements are due to different approaches including improved material quality, the use of tandem cells with higher emitter temperatures, and near-perfect sub-bandgap reflectance. Tervo et al.[14] in 2022 demonstrated a single junction 0.74 eV GaInAs thermophotovoltaic device reaching an efficiency of 38.8% and an electrical power density of 3.78 W/cm$^2$ at an emitter temperature of 1850 °C. Achieving such a high power density and efficiency at such a high emitter temperature is the result of a combination of effective spectral management, an optimized cell architecture, superior material quality, and minimal series resistance. In the same year, LaPotin et al.[15] fabricated two-junction 1.4/1.2 eV GaAs/GaInAs tandem devices using an inverted metamorphic multijunction architecture[16–18]. They reported an efficiency of 41.1% while operating at a power density of 2.39 W/cm$^2$ and an emitter temperature of 2400 °C. Multi-junction cells increase efficiency over single junctions by reducing thermalization losses and reducing resistive losses by operating at lower current density. In 2024, Roy-Layinde et al.[19] achieved a record power conversion efficiency of 44% for a 0.9 eV GaInAsP air-bridge cell at 1435 °C. The air-bridge design involves using a thin-film TPV membrane supported by gold gridlines, which creates a gap between the semiconductor material and the back reflective metal, leading to near-unity sub-bandgap reflectance[20,21]. This structure helps optimize the recycling of low energy photons back to the thermal emitter.

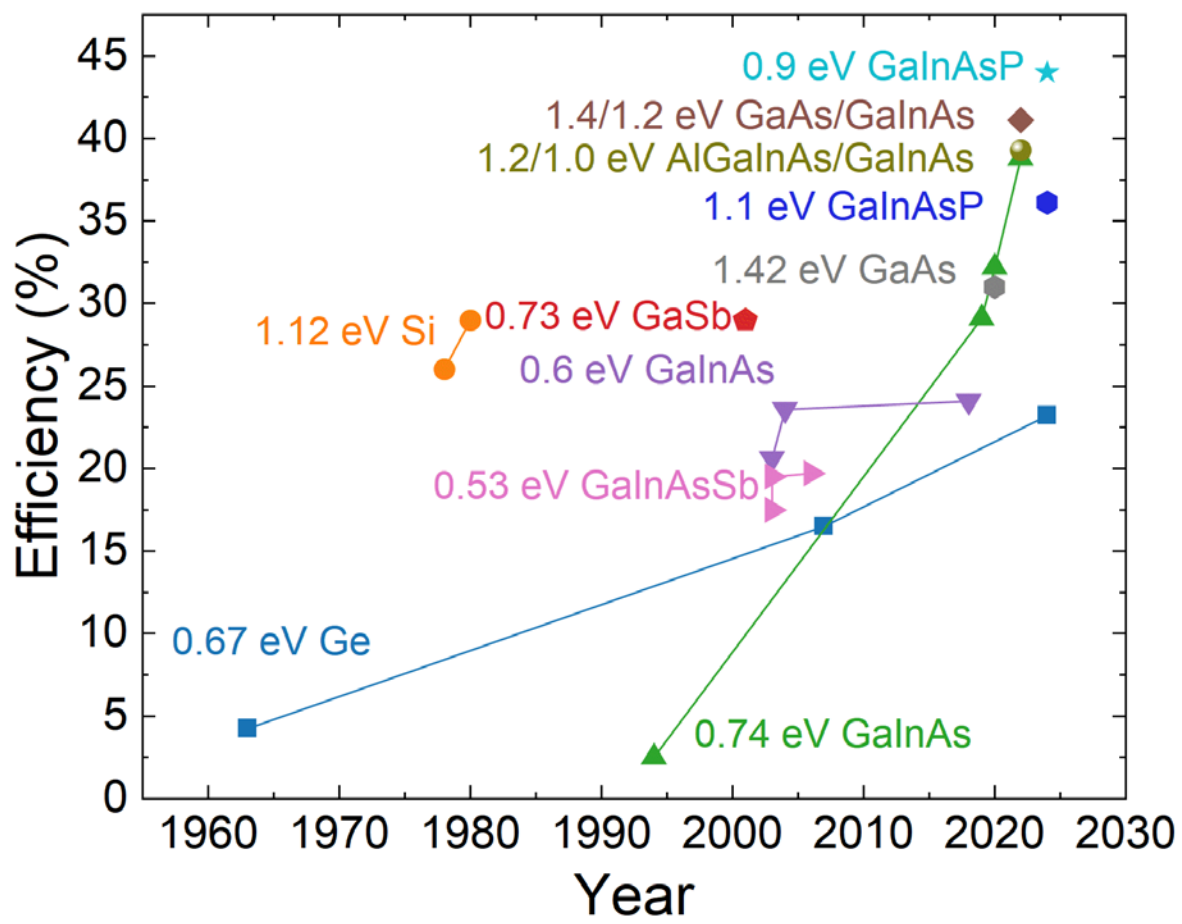

**Figure 2.** Historical best measured single junction and multijunction TPV efficiencies for different cell types. Reporting literature: 0.53 eV GaInAsSb [22–24], 0.6 eV GaInAs [25–28], 0.67 eV Ge [29–31], 0.73 eV GaSb [32], 0.74e V GaInAs [14,33,34], 1.12 eV Si[35,36], 1.4/1.2 eV GaAs/GaInAs [15], 1.2/1.0eV AlGaInAs/GaInAs[15], 0.9 eV GaInAsP[19], 1.1 eV GaInAsP[19], 1.42 eV GaAs[37]. Data is provided in the supplementary material.

Although there have been rapid recent improvements in efficiency, the various approaches have resulted in inconsistencies in the trends for efficiency with respect to bandgap energy and emitter temperature, as shown in Fig. 3. At mid-range temperatures (1000 °C – 2000 °C), for example, intermediate bandgaps of 0.73 or 0.74 eV have resulted in both some of the highest and lowest efficiencies. 0.53 eV cells have performed better than 0.6 eV cells but worse than a 0.9 eV cell. Furthermore, many of the highest-efficiency results are in this temperature range, despite recent experiments with substantially higher emitter temperatures that should in principle perform better. Finally, lower efficiencies than expected were recorded for 0.35 eV bandgap cells at lower temperatures. For example, our model would predict an efficiency greater than 9% for an InAs cell with emitter temperature between 500 °C – 1000 °C as we will discuss later. The lack of consistency makes it hard to predict the best bandgap for a certain temperature or application and further shows that factors other than bandgap play a significant role.

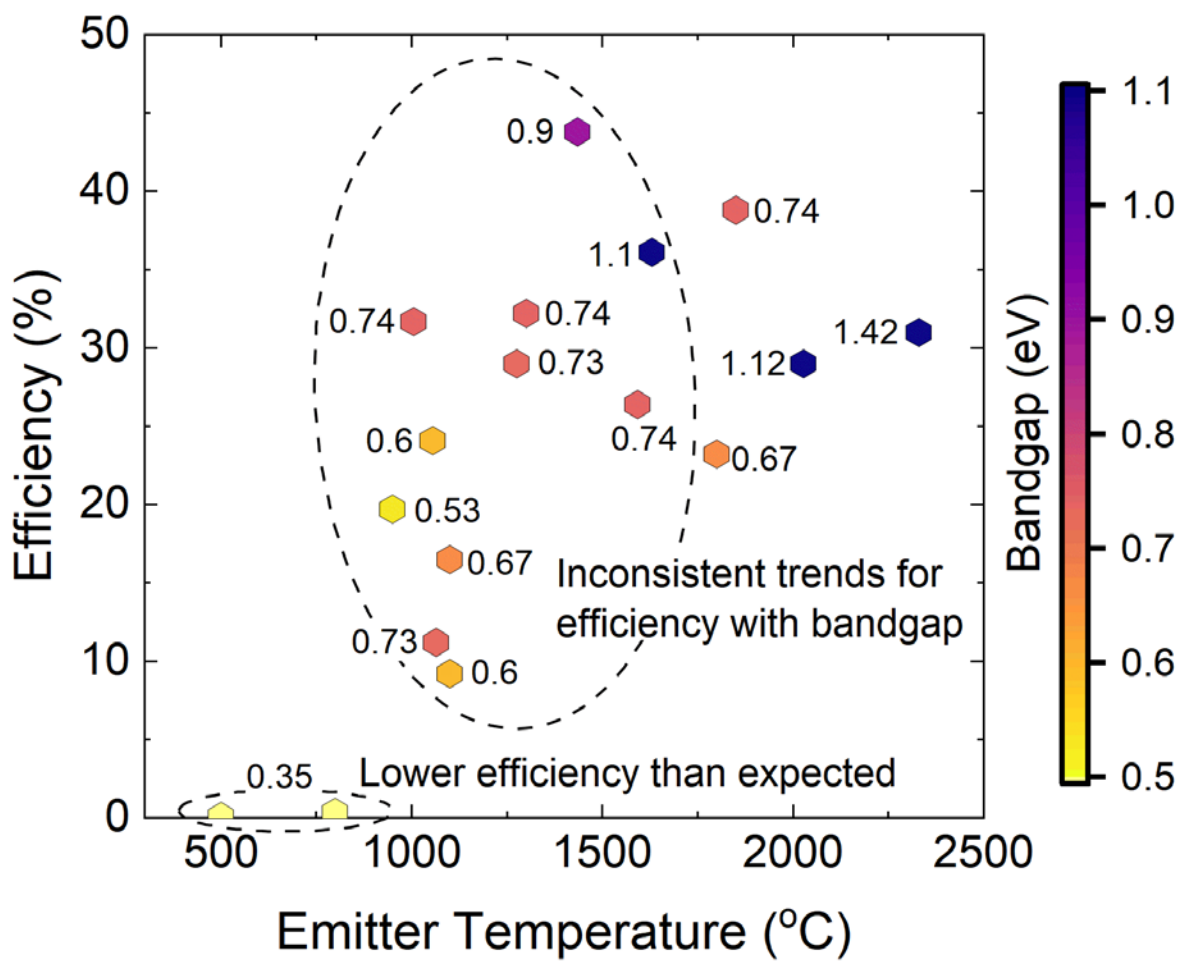

**Figure 3.** A plot of historical TPV efficiencies showing inconsistencies in the trends for efficiency with bandgap energy. Literature: 0.35 eV[38], 0.53 eV[22], 0.6 eV[26,28], 0.67 eV [30,31], 0.73 eV[32,39], 0.74 eV[14,37,40,41], 0.9 eV[19], 1.1 eV[19], 1.12 eV[36], 1.42 eV[37]. Data is provided in the supplementary material.

To better understand differences in past results and allow researchers to reliably predict the performance of TPV systems with varying bandgaps, emitter temperatures, and other characteristics in a variety of applications, we present here an empirical model for TPV performance. We first describe the model and its dependence on the eight parameters mentioned previously: emitter temperature, bandgap energy, view factor, above and below-bandgap (effective) emittance, internal quantum efficiency, saturation current, and series resistance. Importantly, we show how measured or expected open-circuit voltage (and the associated bandgap-voltage offset) can be used as a reliable indication of material quality instead of the saturation current, which is more difficult to predict. We then validate our model against past results, and we use it to show the impacts of different input parameters on device efficiency and power density. Finally, we predict the efficiency and power density across a range of emitter temperatures and bandgap energies considering different important effects like cell quality and sub-bandgap reflectance.

## METHODS

The efficiency of a TPV system is the product of the transfer efficiency $TE$ from the heat source to the thermal emitter, the cavity efficiency $CE$ that indicates how well the emitter and cell are radiatively integrated and thermally insulated, and the efficiency of the emitter-cell pair $\eta_{TPV}$ [14,42]:

$$\eta_{system} = TE \cdot CE \cdot \eta_{TPV} \tag{1}$$

Here we will refer to $\eta_{TPV}$ as the TPV efficiency, which is the quantity plotted above in Figures 2 and 3, but it is also sometimes referred to as the "pairwise efficiency"[3]. This is defined as the ratio of the electrical output power $P$ to the total radiation absorbed by the cell $Q_{abs}$ [27,35]:

$$\eta_{TPV} = \frac{P}{Q_{abs}} \tag{2}$$

The total radiation absorbed by the cell is

$$Q_{abs} = VF \cdot \left[ \varepsilon_{bel} \int_0^{E_g} Q_{bb}(E, T_{em}) dE + \varepsilon_{abv} \int_{E_g}^{\infty} Q_{bb}(E, T_{em}) dE \right] \quad (3)$$

where $VF$ is the view factor, $E_g$ is the bandgap energy, $Q_{bb}(E, T_{em})$ is the spectral blackbody radiation flux from the emitter at temperature $T_{em}$, and $\varepsilon_{bel}$ and $\varepsilon_{abv}$ are the effective below- and above-bandgap emittance of the emitter-cell pair (this differs from a solar PV where the below bandgap emittance is irrelevant). These account for possible non-unity emittance of both the emitter $\varepsilon_{em}$ and the cell $\varepsilon_{cell}$, and they can be written as[43]

$$\varepsilon_{bel(abv)} = \frac{1}{\frac{1}{\varepsilon_{em,bel(abv)}} + \frac{1}{\varepsilon_{cell,bel(abv)}} - 1} \quad (4)$$

for two infinite, parallel surfaces. This equation can be written for both the above-bandgap and below-bandgap spectral regimes, where $\varepsilon_{em,bel(abv)}$ and $\varepsilon_{cell,bel(abv)}$ are the blackbody-weighted emittances for each spectral regime at the corresponding emitter or cell temperature. It is important to note that an accurate form of Eqn. (3) for low view factors, complex cavity or scattering effects, or cell temperature close to the emitter temperature may be more complicated, but this equation works well for high view factors (as expected in most practical applications), near-black emitters (as used in most TPV experiments), and emitter temperatures much higher than the cell temperature. The effective emittances could also be written in terms of the spectral emittances (inside the integrals over photon energies) of both the emitter and the cell, but the form used here is more convenient when approximating quantities like the weighted above- or below-bandgap reflectance. For example, a black emitter paired with a cell that has a weighted below-bandgap reflectance of 97% would have a corresponding effective below-bandgap emittance (one minus reflectance) of 3%.

The electrical power density is described by the current-voltage characteristics under illumination and can be expressed as the product of the current density $J$ and the voltage $V$. These two quantities are related by the ideal-diode equation[44]

$$J = -J_{sc} + J_0 \left\{ exp\left[ \frac{q(V - J \cdot R_{series})}{kT_{cell}} \right] - 1 \right\} \quad (5)$$

where $J_{sc}$ is the short-circuit current density, $J_0$ is the saturation current density, $R_{series}$ is the series resistance, $T_{cell}$ is the temperature of the cell, and $k$ is Boltzmann's constant. Here we use a single diode with ideality factor $n = 1$ rather than a double diode equation, which we show later to be sufficient, we neglect shunt pathways as high-performance cells should have a very high shunt resistance, and we assume $J_{sc}$ is the same as the photocurrent, which holds for good cells with low series resistance. $J_{sc}$ can be obtained from a similar equation to Eqn. (3) considering only the above-bandgap radiation and dividing by photon energy to obtain the number of absorbed photons. This is multiplied by the electron charge $q$ to obtain current and the internal quantum efficiency $IQE$ to account for nonidealities in charge collection. Although $IQE$ is usually reported on a spectral basis, TPV cells typically have a fairly flat, high region of $IQE$ above the bandgap energy that covers most of the above-bandgap radiation from the thermal emitter[14,15,28,38]. $J_{sc}$ can therefore be written as

$$J_{sc} = q \cdot VF \cdot IQE \cdot \varepsilon_{abv} \int_{E_g}^{\infty} \frac{1}{E} Q_{bb}(E,T)\, dE \tag{6}$$

If desired, the spectral values of $\varepsilon_{abv} \cdot IQE$ (equal to external quantum efficiency $EQE$) could also be used in Eqn. (6) by moving that term inside the integral. We can now point out that, in principle, the previous equations and inputs comprise all that is needed to predict the performance of an idealized TPV system. If one knows the temperature of the emitter, the reflectance or emittance of the emitter and cell ($\varepsilon_{bel}$ and $\varepsilon_{abv}$), the view factor, and the cell characteristics $E_g$, $IQE$, $R_{series}$, $J_0$, and $T_{cell}$, then the previous equations can be solved for the radiation absorbed, the current-voltage characteristics (which provide power density), and the efficiency. $T_{em}$, $T_{cell}$, and $VF$ are application dependent, although $VF$ is likely to be near unity in most practical applications. $E_g$ is set by the cell materials, although the proper choice of materials for a particular system remains unclear, as discussed earlier. $\varepsilon_{bel}$ and $\varepsilon_{abv}$ depend on the optical properties of the emitter and cell, although recent works have tended to use cells with sub-bandgap reflectance greater than 92% ($\varepsilon_{bel} < 8\%$)[14,15,40,45]. $IQE$ can vary, but for good cells this is nearly always greater than 95% across a wide range of bandgaps[14,15,17,28] . The series resistance is typically in the range of 5 – 1,000 mΩ cm$^2$ and is most often below 50 mΩ cm$^2$ [14,15,19] for good cells . That leaves just the saturation current, which is more difficult to predict.

A look at past literature for values of $J_0$ across bandgap energies and material quality show wide variations, as indicated in Fig. 4a for the cells reported in the previous figures and for a number of additional lattice-matched and lattice-mismatched III-V cells fabricated by some of us at the National Renewable Energy Laboratory over many years. It has the expected negative exponential dependence on bandgap energy, but it also varies by orders of magnitude for a single bandgap energy (e.g., $10^{-14}$ to $10^{-10}$ A/cm$^2$ for 1.1 eV Si or $10^{-10}$ to $10^{-7}$ A/cm$^2$ for 0.74 eV GaInAs). This variation makes it challenging to compare saturation current values between different cells or predict an expected value. A different, related quantity that is more familiar and easier to predict is the open-circuit voltage $V_{oc}$. For a particular $J_{sc}$ and $T_{cell}$, the diode equation can be rearranged to express $V_{oc}$ as

$$\mathrm{V_{oc}} = \frac{\mathrm{kT_{cell}}}{\mathrm{q}} \ln\left(\frac{\mathrm{J_{sc}}}{\mathrm{J_0}} + 1\right) \tag{7}$$

$V_{oc}$ is still strongly dependent on the bandgap energy through the $J_0$ term, but this can be transformed into a more consistent and insightful metric: the bandgap-voltage offset $W_{oc}$, which is defined as the difference between the bandgap voltage and the open-circuit voltage of the cell[46]:

$$\mathrm{W_{oc}} = \frac{\mathrm{E_g}}{\mathrm{q}} - \mathrm{V_{oc}} \tag{8}$$

The bandgap-voltage offset depends only weakly on the bandgap, which results in a nearly constant value for many different materials and bandgap energies. A plot of values of $W_{oc}$ for different bandgaps determined from the data in Fig. 4a assuming a constant short-circuit current density of 15 mA/cm$^2$ is given in Fig. 4b. These calculations use Eqn. (7) with $J_{sc}$ = 15 mA/cm$^2$, $T_{cell} = 25$ °C, and $J_0$ from Fig. 4a to determine $V_{oc}$, then use Eqn. (8) with the bandgap energy to determine $W_{oc}$. The values for all of these quantities are provided in the supplementary material.

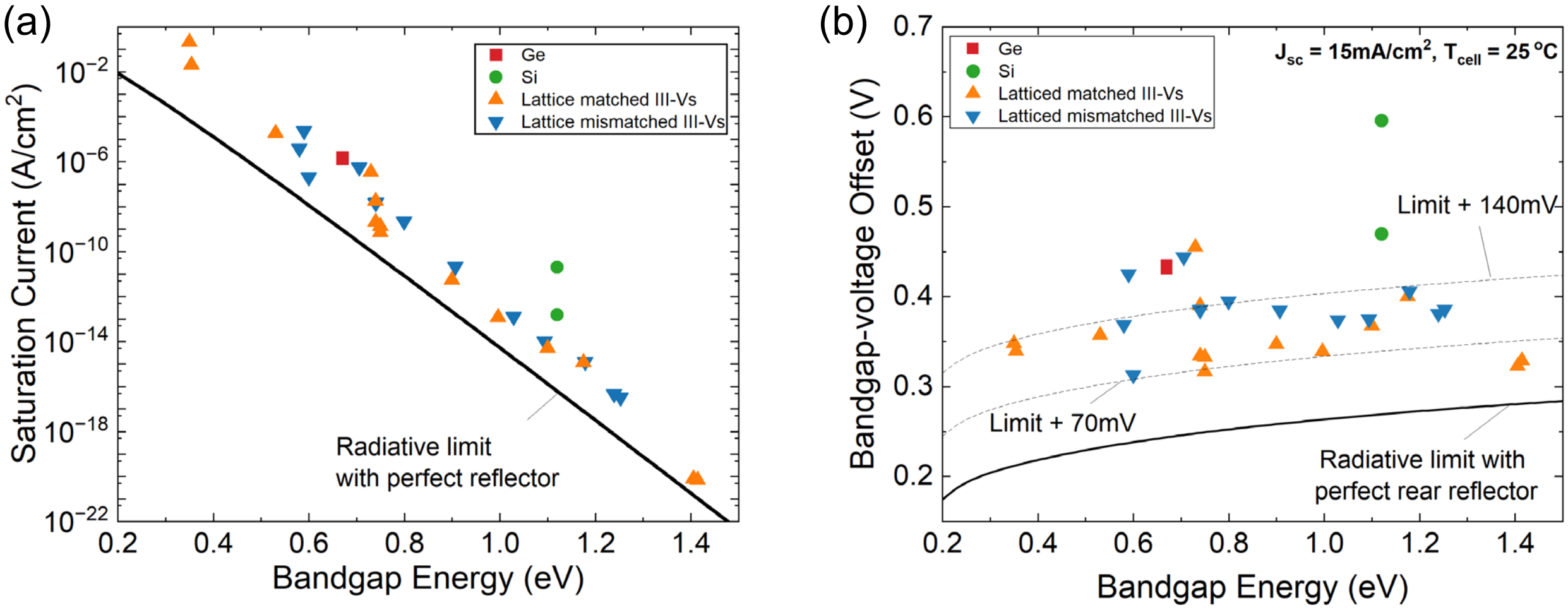

**Figure 4.** (a) Plot of saturation current as a function of bandgap energy for various materials. (b) Plot of bandgap-voltage offset ($W_{oc}$) as a function of bandgap energy calculated from (a). Data corresponds to references in Figures 2 and 3, InAs[38,47], as well as NREL-fabricated cells[48–50]. Data is provided in the supplementary material.

For III-V cells, a "good" $W_{oc}$ typically falls within the range of 0.3 V to 0.4 V, indicating efficient cell operation with minimal losses, while less optimal III-V $W_{oc}$ values exceed 0.4 V, corresponding to greater voltage losses. To better capture the slight dependence on bandgap energy, we add 70 mV and 140 mV to the radiative limit, and this range includes most of the III-V devices which perform well. These bounding values therefore represent practically attainable good performance for both lattice matched III-Vs, and lattice-mismatched III-Vs. The limit plus 70 mV would represent excellent material quality, and the limit plus 140 mV would represent good material quality. The indirect bandgap Si and Ge TPV cells unsurprisingly show larger $W_{oc}$ values outside of the range identified for III-Vs. However, these values can be improved. A Si heterostructure cell designed for solar applications, for instance, achieved a 15 mA/cm$^2$ $W_{oc}$ of 0.395 V[51], which is quite close to the upper range of III-Vs. We also note that more studies are needed for TPV cells with bandgaps less than about 0.5 eV to see how these trends hold for narrow bandgaps. Using our bounds for $W_{oc}$ for III-Vs, we can predict $W_{oc}$ (and therefore $J_0$) for a desired or expected III-V material and quality. Now, the inputs to our model are complete.

Using measured current-voltage data under illumination and known or estimated values for series resistance and effective emittances, our model can predict the performances of cells in a TPV system across various emitter temperatures. Another approach is to target an achievable $W_{oc}$ value along with other expected parameters and calculate the cell performance for different bandgap energies under illumination from an emitter at a particular temperature, as will be shown in the following sections.

## VALIDATION OF MODEL

In order to ensure the reliability and accuracy of our model in predicting the performance of TPV systems, we performed a validation against experimental data for TPV cells reported by Tervo et al. (0.74 eV GaInAs)[14], Roy-Layinde et al. (0.9 eV InGaAsP) [19], Fernández et al. (0.67 eV Ge) [30] and Lee et al. (1.12 eV Si) [45]. Inputs to the model were series resistance, above and below bandgap reflectance, view factor, bandgap, together with the open circuit voltage, short circuit current, and fill factor at some illumination, and we modelled the emitter as a blackbody. All cells were assumed to be at 25 °C. These values were used in calculating $W_{oc}$ and then the saturation current. The model's results for efficiency and power density were then compared to experimental results as reported by the authors.

Tervo et al. reported the design, fabrication and testing of a lattice-matched 0.74 eV GaInAs cell reaching an efficiency of 38.8% and an electrical power density of 3.78 W/cm$^2$ at an emitter temperature of 1850 °C, with open-circuit voltage of 0.569 V, short-circuit current density of 8.66 A/cm$^2$, series resistance of 6.5 mΩ cm$^2$, above-bandgap emittance of 60%, sub bandgap reflectance of 94.7%, and view factor of 0.31. This corresponds to a $W_{oc}$ of 334 mV at 15 mA/cm$^2$ (radiative limit plus 85 mV), representing a high material quality.

We also considered a lattice-matched 0.9 eV GaInAsP airbridge TPV cell fabricated by Roy-Layinde et al[19]. From communication with the authors, we obtained a sub-bandgap reflectance of 98.8%. Other inputs to the model were directly taken from this paper including an IQE of 98%, an open-circuit voltage of 0.635 V and short-circuit current density of 0.139 A/cm$^2$ at an emitter temperature of 1001°C, series resistance of 40 mΩ cm$^2$, a shunt resistance of 825 kΩ cm$^2$, and a view factor of 0.37. The calculated $W_{oc}$ of this cell at 15 mA/cm$^2$ is 347 mV (radiative limit plus 83 mV), consistent with excellent material quality. This represents a case of a very high sub-bandgap reflectance (very low $\varepsilon_{bel}$).

Data for a 1.1 eV silicon airbridge TPV cell fabricated by Lee et al.[45] was used as well. Inputs taken directly from this paper include a short-circuit current density of 0.922 A/cm$^2$ and open-circuit voltage of 0.63 V at an emitter temperature of 1715 °C and view factor of 0.29. The above-bandgap reflectance is 83.5%, and the series resistance is 9.7 Ω.cm$^2$. A sub-bandgap reflectance of 98.5% was inferred from a plot in the paper. This cell has a corresponding $W_{oc}$ of 596 mV (radiative limit plus 327 mV).

The final data input to validate our model was the 0.67 eV Germanium TPV cell fabricated by Fernández et al[30]. Inputs to the model include a short-circuit current density of 1.65 A/cm$^2$ and an open-circuit voltage of 0.353 V at an emitter temperature of 1100 °C and view factor of 1. An above bandgap emittance of 90% was inferred from the plotted external quantum efficiency. These data yield a $W_{oc}$ of 435 mV at 15 mA/cm$^2$ (radiative limit plus 192 mV). The Ge and Si cells are useful reference cases for indirect bandgap materials.

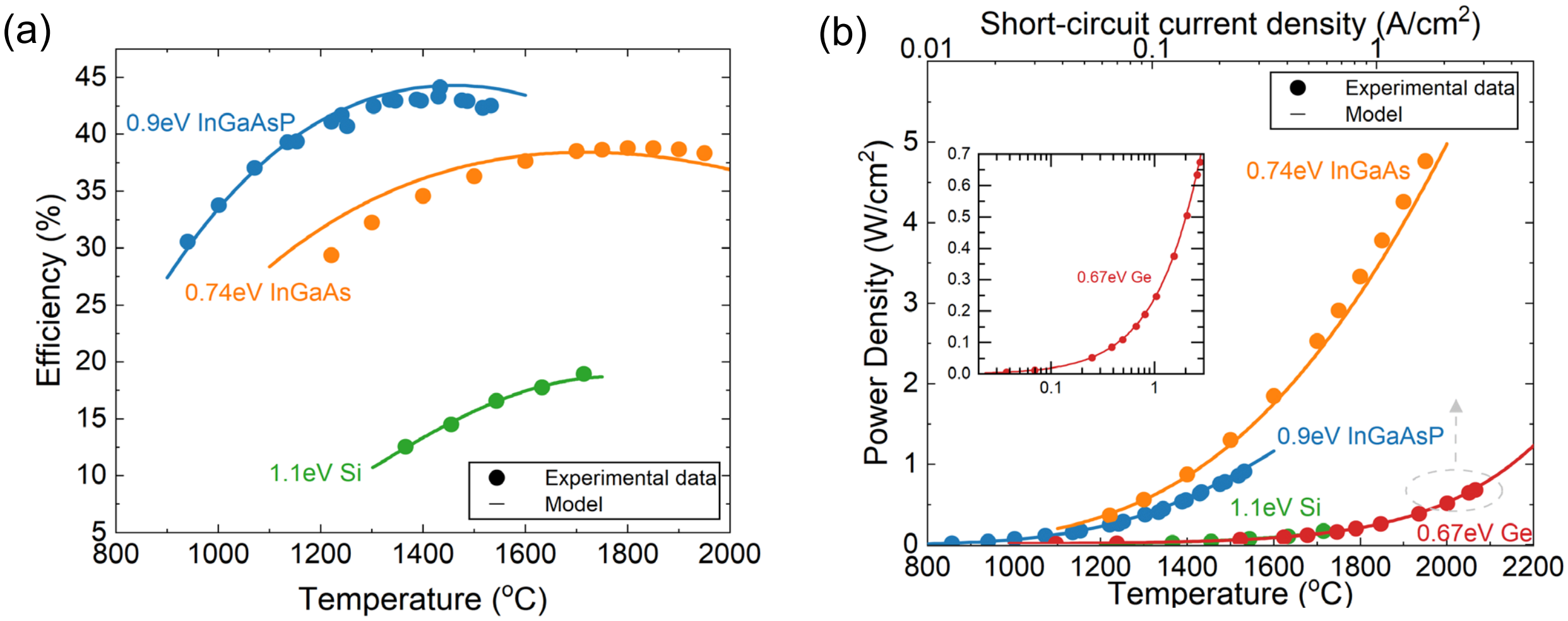

**Figure 5.** Validation of our modeled (a) efficiency and (b) power density as a function of emitter temperature. The points and curve for Ge in (b) are plotted against the top current axis due to data availability, while the other data are plotted against the bottom temperature axis.

Using the inputs provided above, we used our model to generate predicted efficiency and power density and compared the model with the experimental data as reported by the authors. The validation of the power density model for 0.67 eV Germanium was plotted against short-circuit current density instead of emitter temperature to match the experimental data from the authors, and no efficiency data was available for this device. The results of the comparison are presented in Fig. 5, and we observe an excellent agreement between our model and the experiments. This provides confidence in the use of the model to predict TPV performance.

## RESULTS AND DISCUSSION

### Importance of considering spectral and electrical losses in TPV predictions

In thermophotovoltaic systems, accurately predicting the performance is important for optimizing efficiency and effectiveness. Spectral and electrical losses are two key factors that significantly impact the overall performance of TPV systems. Spectral losses refer to the losses that arise due to a mismatch between the emitted thermal radiation spectrum and TPV cell's absorption characteristics, and in our model, these are captured by the $\varepsilon_{abv}$ and $\varepsilon_{bel}$ terms. Electrical losses refer to the losses that occur during the conversion of the absorbed photon into electrical energy, and these are accounted for by the other model parameters. Datas et al[52] presented the optimum semiconductor bandgaps leading to the maximum efficiency and power density in TPV converters by considering only spectral losses. In Fig. 6a and b, we show the impact of both spectral and electrical losses on TPV efficiency and power density for various bandgap energies with a fixed emitter temperature of 1400°C. For these results, we use $\varepsilon_{bel} = 0.1$, $\varepsilon_{abv} = 1$, $VF = 1$, $IQE = 100\%$, $W_{oc} = 0.35$ (corresponding to a decent III-V cell), and $R_{series} = 5$ mΩ cm$^2$. From figure 6a, we observe that for this emitter temperature and bandgaps lower than about 1.2eV, considering only spectral losses results in an overestimation of the efficiency of the TPV system while higher bandgaps (above 1.2 eV) are not significantly affected. The power density of the system is not changed from the radiative limit by the introduction of spectral losses only (Fig. 6b), because in this case we have kept $\varepsilon_{abv} = 1$ and $VF = 1$, and using $\varepsilon_{bel} < 1$ does not affect power output. The introduction of electrical losses does significantly reduce the power density for bandgaps below about 1 eV, though. This is due both to the higher current density at lower bandgaps driving a higher series resistance power loss, and the difference between the voltage and its radiative limit being proportionally greater at lower bandgaps.

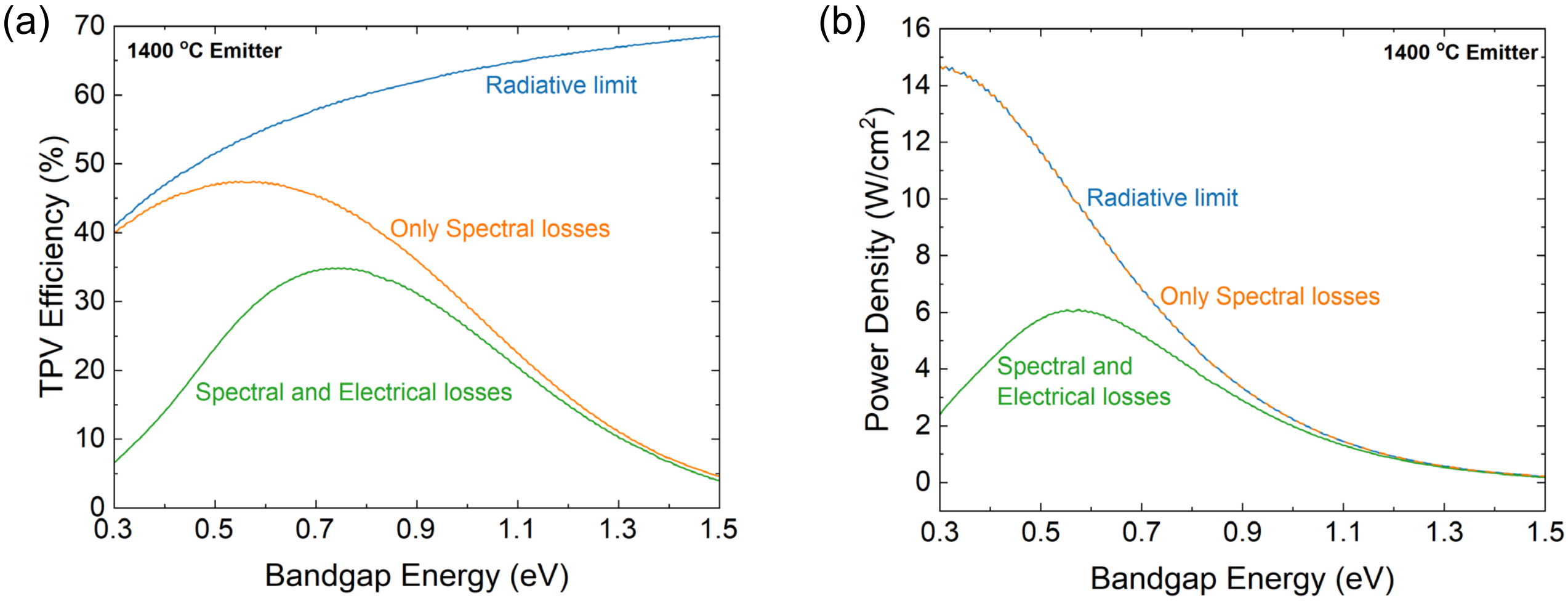

**Figure 6:** (a) Plot of TPV efficiency as a function of bandgap energy showing the radiative limit, only spectral losses, and both spectral and electrical losses. (b) Plot of power density as a function of bandgap energy in the radiative limit and when spectral and electrical losses are considered. The radiative limit and only spectral losses are coincident in (b), so these are shown as an alternating dashed orange and blue line.

## Influence of different model parameters on efficiency and power density:

To gain insight on what strongly affects efficiency and power density, we varied some model parameters from a baseline case. This baseline was the 0.74 eV GaInAs cell fabricated by Tervo et al. (2022)[14] with a series resistance of 6.5 mΩ cm$^2$, open circuit voltage of 0.569V at 1850ºC ($W_{oc} = 0.334$ V at 15 mA/cm$^2$), 40% above bandgap reflectance, 95% below bandgap reflectance, $IQE$ of 100%, $T_{cell} = 25$ °C, and view factor of 0.31. The effects of modifying some of these parameters on the efficiency and power density are shown in Figures 7a and b.

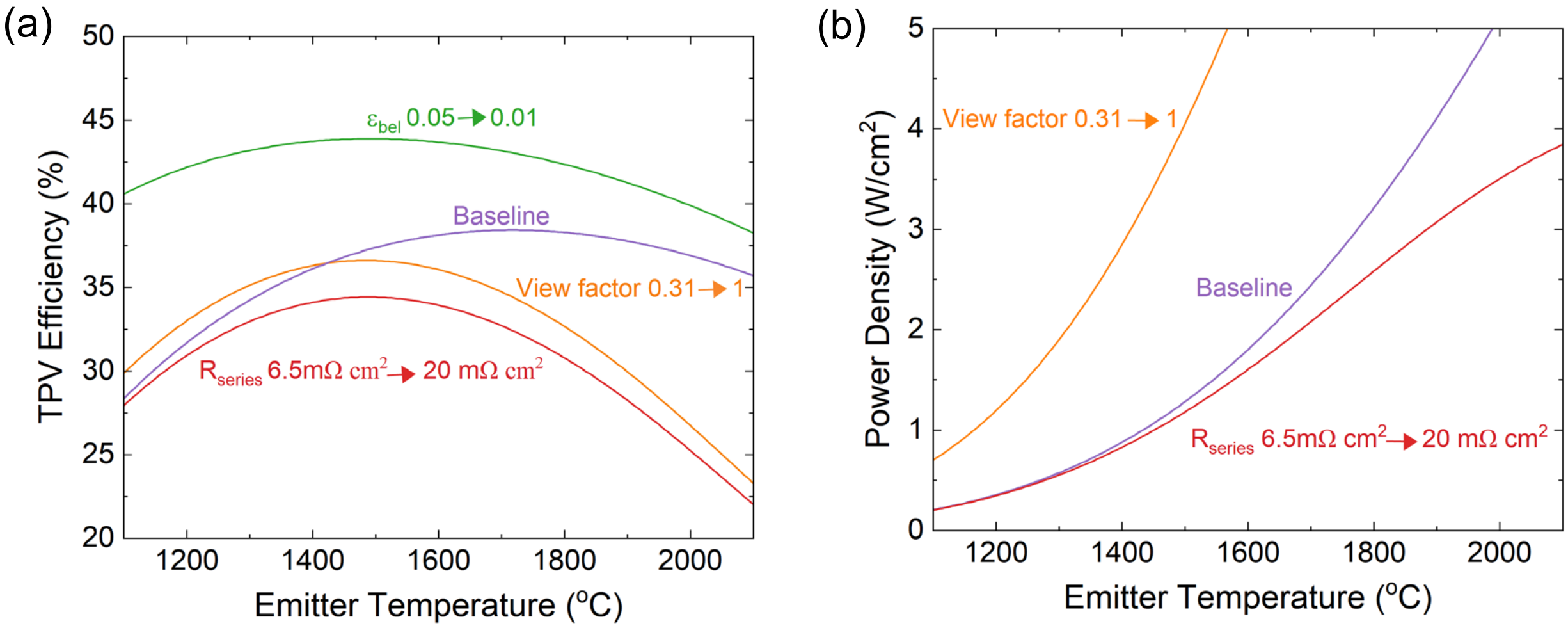

**Figure 7.** Effects of changes in model parameters on TPV (a) efficiency and (b) power density as a function of emitter temperature.

To demonstrate the impact of series resistance, we increased the series resistance from 6.5 mΩ cm$^2$ by a modest amount to 20 mΩ cm$^2$. Increasing the series resistance reduces both the efficiency and power density for all emitter temperatures. This effect is very significant when the irradiance is high, because resistive losses scale with the square of the current, which highlights the importance of minimizing series resistance when the cell is used with very high emitter temperatures. We also examined the effect of improved sub-bandgap reflectance by increasing this from 94.7% to 99%, which means $\varepsilon_{bel}$ decreases from 0.05 to 0.01. As has been shown by other authors[40,45,53] and is shown in Fig. 7a, very high sub-bandgap reflectance can lead to much higher efficiency for lower emitter temperatures, possibly motivating the use of wider bandgap cells in this scenario. This has no impact on power density (no change from the baseline case in Fig. 7b), as the above-bandgap characteristics are unchanged. Finally, we looked at the effect of increased view factor on the baseline result, which would be more representative of a practical system, by increasing this from 0.31 to 1. At lower temperatures (below 1400 °C) an increased view factor led to a slight increase in efficiency due to the logarithmic increase in voltage with current density (see Eq. 7), but at higher temperatures the efficiency is significantly reduced as series resistance becomes increasingly dominant with increasing current densities. In many applications, this higher power density may be a favorable tradeoff for reduced efficiency.

## Practical predictions of TPV performance

The most useful aspect of our model is to predict the trends in efficiency and power density for different bandgaps, emitter temperatures, and cell characteristics. This can allow a designer to select a bandgap or material, especially when considering III-V cells where bandgap may be tuned by composition, for a particular application based on desired performance metrics. Figure 8 shows these performance predictions for both efficiency and power density as a function of emitter temperature for bandgap energies ranging from 0.4 to 0.8 eV. For these calculations, we use $VF = 1$, $\varepsilon_{abv} = 1$, $IQE = 1$, $T_{cell} = 25$ °C, and $R_{series} = 0.01$ Ω cm$^2$. A range of III-V material qualities are indicated by using the previously described boundaries on $W_{oc}$ from the radiative limit plus 70 mV (excellent III-V material quality, top solid lines) to the radiative limit plus 140 mV (good III-V material quality, bottom dashed lines). For efficiency, we use an effective sub-bandgap reflectance of 95% in Fig. 8(a) and 99% in Fig. 8(b), but this does not affect power density in Fig. 8(c).

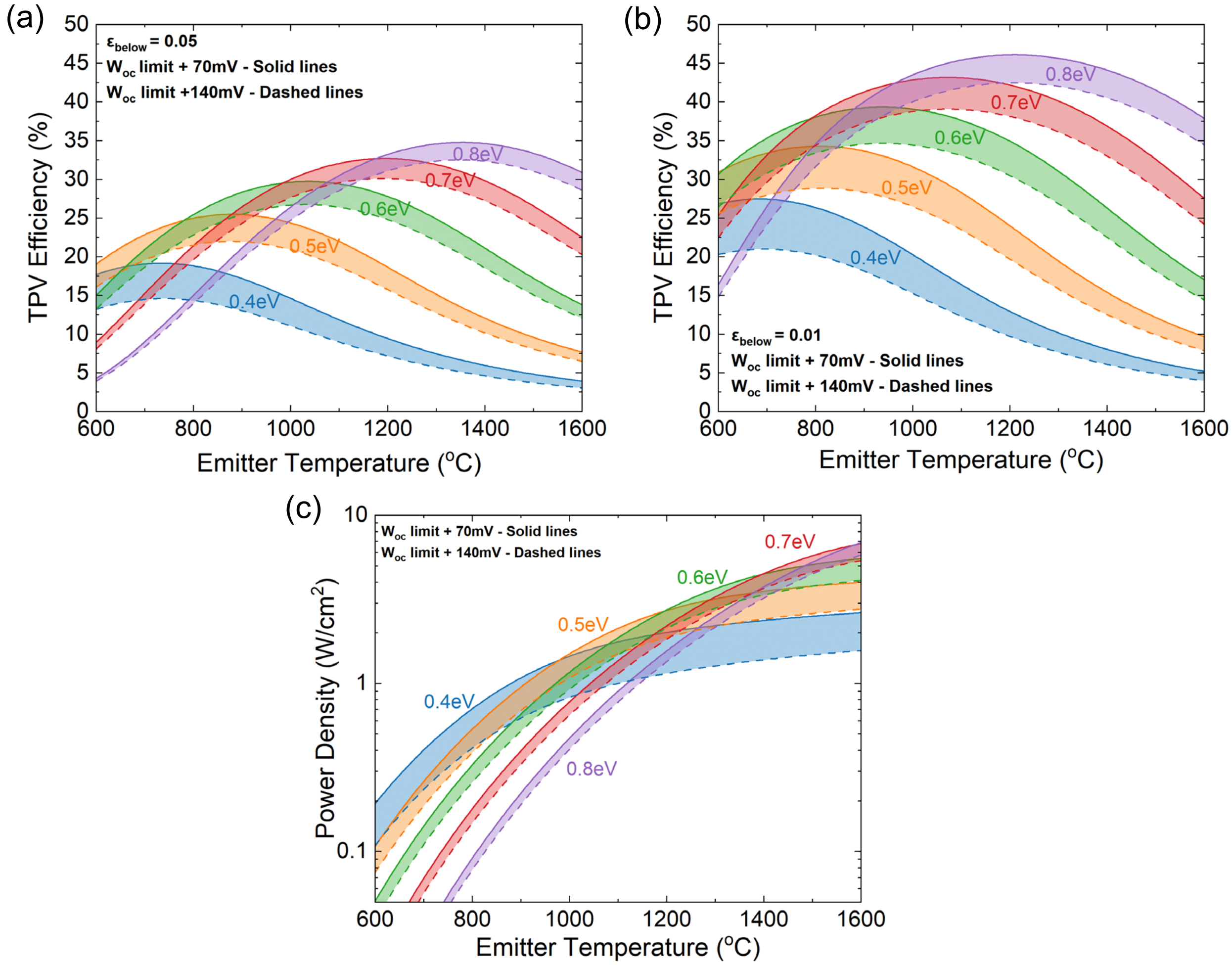

**Figure 8.** TPV (a, b) efficiency and (c) power density for varying emitter temperature and several bandgap energies. The efficiencies are shown for a sub-bandgap reflectance of (a) 95% and (b) 99%; these values do not affect power density in (c). In all panels the top solid lines correspond to $W_{oc}$ limit + 70mV (excellent III-V material quality), and the bottom dashed lines correspond to $W_{oc}$ limit + 140 mV (good III-V material quality).

There are several expected trends in Figure 8 that are worth mentioning. For efficiency, each bandgap energy reaches a peak at some emitter temperature where the losses associated with sub-bandgap absorption, high-energy thermalization, and series resistance are collectively minimized. This peak in efficiency occurs at higher temperatures and reaches higher values for wider bandgap materials. The "best" bandgap energy for operation at a particular temperature is not necessarily the peak efficiency point for a particular bandgap energy. For example, if the cell has $\varepsilon_{bel} = 0.01$ and $W_{oc}$ = limit + 140 mV (Fig. 8b) and the emitter is at 1,000 °C, the optimum bandgap energy is near 0.8 eV, even though a 0.8 eV cell would perform even better at 1,200 °C. For power density, narrower bandgap materials always provide more power for low emitter temperatures ($\lesssim$ 900 °C), as these have more of the thermal spectrum above the bandgap energy. The opposite is true for very high temperatures (~1,600 °C), where wider bandgap materials can more effectively convert the higher energy photons to electricity.

When examining the effects of material quality and sub-bandgap reflectance, there are two important observations from Fig. 8. First, we note that a reduction in material quality (moving from $W_{oc}$ limit + 70 mV to $W_{oc}$ limit + 140 mV) results in a decrease in both efficiency and power density across all conditions, but it does not have a significant effect on the performance trends, peak locations, nor crossover points between curves for different bandgap energies. This means that the optimum choice of bandgap for some application, based on either efficiency or power density, will not change much for different material quality as long as the material quality is not very poor. Second, by increasing the sub-bandgap reflectance (decreasing $\varepsilon_{bel}$), not only does efficiency increase significantly, but the peak efficiency points shift to lower emitter temperatures. As has been pointed out by other authors[40,45,53], this means that very low $\varepsilon_{bel}$ can enable high-efficiency performance with wide-bandgap materials at relatively low emitter temperatures, which may be appealing since wider bandgap materials tend have better performance and are easier to fabricate.

A final important takeaway from Fig. 8 is that very different III-V bandgap energies are preferred when optimizing for efficiency compared to when optimizing for power density[12]. Continuing with the prior example, if the cell has $\varepsilon_{bel} = 0.01$ and $W_{oc} =$ limit + 140 mV (Fig. 8b) and the emitter is at 1,000 °C, the optimum bandgap energy for efficiency is near 0.8 eV. For maximum power density, however, the optimum bandgap energy for the same case (Fig. 8c) is 0.5 eV, a significant difference. The 0.8 eV cell would produce 0.433 W/cm$^2$ in these conditions, whereas the 0.5 eV cell would produce 1.26 W/cm$^2$. These differences are even more significant for lower emitter temperatures: at 700 °C in Fig. 8c, the 0.4 eV cell produces about an order of magnitude more power than the 0.8 eV cell. Because many potential applications are sensitive to system weight, capital cost, or system size, power density may play an equal or even more important role than efficiency in system design. This motivates continued research and development on narrow bandgap III-V TPV cells, whose measured performance has not reached our empirical predictions.

## CONCLUSIONS

In conclusion, using an empirical model based on past cell measurements, we have been able to show realistic performance predictions for various types of cells over a broad range of emitter temperatures. Past work on optimal bandgaps has focused on only spectral losses, but by integrating both spectral and electrical characteristics into our model, we have provided a more useful prediction of TPV performance in real-world scenarios. Our results show that neglecting electrical losses can lead to overestimation of system performance. We also demonstrated the influence of the series resistance, view factor and sub-bandgap reflectance on TPV performance. Our results showed that minimizing series resistance while maximizing sub-bandgap reflectance is an effective path to increase the efficiency of the system. Generally, an increase in view factor results in a significant increase in the power density of the system.

Finally, we predicted the best III-V bandgaps for maximum efficiency or power density over a wide range of temperatures by considering cell material quality and sub-bandgap reflectance. Increasing sub-bandgap reflectance will lead to a significant increase in the efficiency of TPV cells and could allow wider bandgap materials to exhibit good efficiency at lower emitter temperatures. On the other hand, we demonstrated that a significant trade-off exists between power density and efficiency, with lower bandgaps preferred for high power density. Certain applications, such as waste heat recovery or space-based power generation, may prioritize power density over efficiency. This trade-off underscores the necessity of application-specific optimization during TPV design.

**ACKNOWLEDGEMENTS**

T.M.D. and E.J.T. acknowledge support by the University of Wisconsin–Madison Office of the Vice Chancellor for Research with funding from the Wisconsin Alumni Research Foundation. This work was authored in part by the National Renewable Energy Laboratory, operated by Alliance for Sustainable Energy, LLC, for the U.S. Department of Energy (DOE) under Contract No. DE-AC36-08GO28308. The views expressed in the article do not necessarily represent the views of the DOE or the U.S. Government. The U.S. Government retains and the publisher, by accepting the article for publication, acknowledges that the U.S. Government retains a nonexclusive, paid-up, irrevocable, worldwide license to publish or reproduce the published form of this work or allow others to do so, for U.S. Government purposes. We thank the two anonymous reviewers for their comments which helped to improve this work.